\documentclass[aps,eqsecnum,preprint,floats,epsf,epsfig,nofootinbib,letter]{revtex4}
\usepackage{epsfig}
\usepackage{multirow}

\begin{document}
\def\be{\begin{eqnarray}}
\def\en{\end{eqnarray}}
\def\non{\nonumber}
\def\la{\langle}
\def\ra{\rangle}
\def\nc{N_c^{\rm eff}}
\def\vp{\varepsilon}
\def\drho{\bar\rho}
\def\deta{\bar\eta}
\def\CP{{\it CP}~}
\def\a{{\cal A}}
\def\B{{\cal B}}
\def\c{{\cal C}}
\def\d{{\cal D}}
\def\e{{\cal E}}
\def\p{{\cal P}}
\def\t{{\cal T}}
\def\up{\uparrow}
\def\dw{\downarrow}
\def\vma{{_{V-A}}}
\def\vpa{{_{V+A}}}
\def\smp{{_{S-P}}}
\def\spp{{_{S+P}}}
\def\J{{J/\psi}}
\def\ov{\overline}
\def\Lqcd{{\Lambda_{\rm QCD}}}
\def\pr{{Phys. Rev.}~}
\def\prl{{Phys. Rev. Lett.}~}
\def\pl{{Phys. Lett.}~}
\def\np{{Nucl. Phys.}~}
\def\zp{{Z. Phys.}~}
\def\lsim{ {\ \lower-1.2pt\vbox{\hbox{\rlap{$<$}\lower5pt\vbox{\hbox{$\sim$}
}}}\ } }
\def\gsim{ {\ \lower-1.2pt\vbox{\hbox{\rlap{$>$}\lower5pt\vbox{\hbox{$\sim$}
}}}\ } }

\font\el=cmbx10 scaled \magstep2{\obeylines\hfill May, 2014}

\vskip 1.5 cm

\centerline{\large\bf Near Mass Degeneracy in the Scalar Meson Sector:}
\centerline{\large\bf Implications for $B^*_{(s)0}$ and $B'_{(s)1}$ Mesons}

\bigskip
\centerline{\bf Hai-Yang Cheng$^{1}$, Fu-Sheng Yu$^{2}$}
\medskip
\centerline{$^1$ Institute of Physics, Academia Sinica}
\centerline{Taipei, Taiwan 115, Republic of China}
\medskip
\centerline{$^2$ School of Nuclear Science and Technology, Lanzhou University} \centerline{Lanzhou 730000, People's Republic of China}
\medskip

\bigskip
\bigskip
\bigskip
\bigskip
\bigskip
\centerline{\bf Abstract}
\bigskip
\small

The empirical observation of near degeneracy of scalar mesons above 1 GeV, namely, the mass of the strange-flavor scalar meson is similar to that of the non-strange one, is at variance with the naive expectation of the quark model.
Qualitatively, the approximate mass degeneracy can be understood as a consequence of self-energy effects due to strong coupled channels which will push down the mass of the heavy scalar meson in the strange sector more than that in the non-strange partner. However, it works in the conventional model without heavy quark expansion, but not in the approach of heavy meson chiral perturbation theory as mass degeneracy and the physical masses of $D_{s0}^*$ and $D_0^*$ cannot be accounted for simultaneously.
In the heavy quark limit, near mass degeneracy observed in the scalar charm sector will imply the same phenomenon in the $B$ system.  We have the prediction $M_{B_0^*}\approx M_{B_{s0}^*}\approx 5715\,{\rm MeV}+\delta\Delta_S$  based on heavy quark symmetry and the leading-order QCD correction, where $\delta\Delta_S$ arises from $1/m_Q$ corrections. A crude estimate indicates that $\delta\Delta_S$ is of order $-35$ MeV or less.
We stress that the closeness of $B_{s0}^*$ and $B_0^*$ masses implied by heavy quark symmetry is not spoiled by $1/m_Q$ or QCD corrections. The mass-shift effect on $K_0^*(1430)$ is discussed.

\pagebreak

\section{Introduction}
Although the hadron properties and spectroscopy have been described successfully by the SU(3) quark model, the empirical observation of the near mass degeneracy of scalar mesons with masses above 1 GeV, namely, the mass of the strange-flavor scalar meson is similar to that of the non-strange one, is at variance with the naive expectation of the quark model where the mass difference is expected to be of order 80$\,\sim$110 MeV.
For example, the mass of $D_0^*(2400)^0$, $2318\pm29$ MeV (see Table \ref{tab:data}), \footnote{This is the average of the masses $2297\pm8\pm20$ MeV by BaBar \cite{BaBar:D0}, $2308\pm17\pm32$ MeV by Belle \cite{Belle:D0} and $2407\pm21\pm35$ MeV by FOCUS \cite{FOCUS:D0}. We will not consider the measurement of $M_{D_0^{*\pm}}=2403\pm14\pm35$ MeV by FOCUS as it will imply that $D_0^{*\pm}$ is heavier than $D_{s0}^*$ (even after taking into account experimental errors) despite a strange quark for the latter.}
is almost identical to the mass, $2317.8\pm0.6$ MeV, of $D_{s0}^*(2317)$, while the quark model predicts $M_{D_{s0}^*}\sim {\cal O}(2480)$ MeV and $M_{D_{0}^*}\sim {\cal O}(2390)$ MeV \cite{Godfrey,Di Pierro}. Also $a_0(1450)$ is almost degenerate with $K_0^*(1430)$ in masses.
For the light scalar mesons, it is naively expected in the 2-quark model that $a_0(980)$ is as light as the $f_0(500)$ (or $\sigma$) meson, whereas experimentally it is degenerate with $f_0(980)$.

The accumulated lattice results also hint at an
SU(3) symmetry in the scalar meson sector. Indeed, the near
degeneracy of $K_0^*(1430)$, $a_0(1450)$, and $f_0(1500)$ implies
that, to first order approximation, flavor SU(3) is a good
symmetry for the scalar mesons above 1 GeV, much better than the
pseudoscalar, vector, axial, and tensor sectors.
A quenched lattice
calculation of the isovector scalar meson $a_0$ mass has been
carried out for a range of low quark masses~\cite{Mathur}. It is
found that, when the quark mass is smaller than that of the
strange, the scalar meson $a_0(1450)$ displays an unusual property of being nearly
independent of the quark mass for quark masses smaller than that of
the strange, in contrast to those of $a_1$ and
other hadrons that have been calculated on the lattice.
The chiral
extrapolated mass $a_0 = 1.42 \pm 0.13$ GeV suggests that
$a_0(1450)$ is a $q\bar{q}$ state.  Furthermore, $K_0^{*}(1430)^+$, an $u\bar{s}$ meson, is
calculated to be $1.41 \pm 0.12$ GeV. This explains the fact that $K_0^{*}(1430)$ is basically
degenerate with $a_0(1450)$ despite having one strange quark. This
unusual behavior is not understood and it serves
as a great challenge to the existing hadron models.

It is worth mentioning that in the work \cite{CCL}, two simple and robust
results have been employed as the
starting point for the mixing model between the isosinglet scalar mesons
$f_0(1710)$, $f_0(1500)$, $f_0(1370)$, and the
glueball. One of them is the approximate flavor SU(3) symmetry in the scalar meson sector above 1 GeV. It has been shown in \cite{CCL} that in the SU(3) symmetry limit,
$f_0(1500)$ becomes a pure SU(3) octet and is degenerate with
$a_0(1450)$, while $f_0(1370)$ is mainly an SU(3) singlet with a
slight mixing with the scalar glueball which is the primary
component of $f_0(1710)$. These features remain essentially
unchanged even when SU(3) breaking is taken into account.

There are several issues with the observation of near mass degeneracy of scalar mesons. First of all, can we understand why the masses of $K_0^*(1430)$ and $D_{s0}^*(2317)$ are very close to their non-strange partners $a_0(1450)$ and $D_0^*(2400)^0$, respectively ? Second, why is $D_{s0}^*(2317)$ anomalously lighter than the naive expectation from the quark model ? Third, will the same phenomenon manifest also in the $B$ meson sector, namely, $B_{s0}^*$ and $B_0^*$, which have not yet been observed, have similar masses ?

There are
some attempts to understand the near degeneracy between $a_0(1450)$ and
$K^*_0(1430)$. One suggestion is to introduce instanton contributions to the QCD sum rules \cite{Jin:light}.
Since the 4-quark light scalar nonet is known to
have a reversed ordering, namely, the scalar strange meson $K_0^*(800)$ (or
$\kappa$) is lighter than the non-strange one such as $f_0(980)$,
it has been proposed in \cite{Schechter} to consider the mixing of
the $q\bar q$ heavy scalar nonet with the light nonet to make
$a_0(1450)$ and $K^*_0(1430)$ closer.  It goes further in
\cite{Maiani} to assume that the observed heavy scalar mesons form
another tetraquark nonet. We note, however, the quenched lattice
calculations in \cite{Mathur}, which presumably gives the bare
$q\bar{q}$ states before mixing with $q^2\bar{q}^2$ via sea quark
loops in the dynamical fermion calculation, already
suggest the near degeneracy between $a_0(1450)$ and
$K^*_0(1430)$.

\begin{table}[t] \label{tab:data}
\centering \caption{Measured masses and widths of some non-charmed scalar mesons and excited charmed mesons \cite{PDG}.}
\begin{tabular}{|c  c c | c c c | }\hline
Meson  &   Mass (MeV)  & $\Gamma$ (MeV) & Meson  & Mass (MeV)\hskip 1 cm  &  $\Gamma$ (MeV)  \\ \hline
$a_0(980)$ & $980\pm20$ & $50-100$ & $f_0(980)$ & $990\pm20$ & $40-100$ \\ \hline
$a_0(1450)$ & $1474\pm19$ & $265\pm13$ & $K_0^*(1430)$ & $1425\pm50$ & $270\pm80$ \\ \hline \hline
$D_0^{*}(2400)^0$ & $2318 \pm 29$ & $267 \pm 40$ & $D_{s0}^{*}(2317)$ & $2317.8 \pm 0.6 $ & $<3.8$\\\hline
$D_0^{*}(2400)^\pm$ & $ 2403 \pm 14 \pm 35$ & $~283 \pm 24 \pm 34~$ & & & \\ \hline
$D_{1}^{\prime  }(2430)^0$ & $ 2427 \pm 26 \pm 25$ & $384^{+107}_{-~75}  \pm 74$ & $D_{s1}^{\prime}(2460)$ & $2459.6 \pm 0.6$ & $<3.5$ \\\hline
  \hline
\end{tabular}
\end{table}

In the quark potential model, the predicted masses for $D_{s0}^*$ and $D_0^*$  are higher than the measured ones by order 160 MeV and 70 MeV, respectively. It is well known that the tetraquark structure for light scalar mesons is a simple mechanism for explaining why $a_0(980)$ is almost degenerate in mass with $f_0(980)$ and why $a_0(980)$ and $f_0(980)$ are narrow in width, while $f_0(500)$ and $K_0^*(800)$ are much broader \cite{Jaffe}. By the same token, it is tempting to argue that the tetraquark picture for scalar charmed mesons can account for the closeness of $D_{s0}^*(2317)$ and $D_0^*(2400)$ masses \cite{Dmitrasinovic}. \footnote{The tetraquark nature for the $D_{s0}^*(2317)$  was first proposed in \cite{Cheng:2003kg}. However, its mass is not the same as the non-strange partner $D_0^*$ unless they are assigned to the ${\bf \bar 3_A}$ multiplet \cite{Dmitrasinovic}.}
It was advocated in \cite{Dmitrasinovic} that the small tetraquark mass of $D_{s0}^*(2317)$ can be explained in terms of the 't Hooft's instanton-induced effective quark interaction.  However, the tetraquark picture for scalar charmed mesons is not supported by lattice and QCD sum rule calculations as discussed below. Indeed, for a multiplet of tetraquark charmed mesons, its 2-quark partner must also exist. Experimentally, neither a {\bf 6}-plet nor a second ${\bf \bar 3}$-plet has been observed.  Indeed,
the non-observation of a heavier and broad $0^+$ $c\bar s$ state is against the tetraquark interpretation of $D_{s0}^*(2317)$.

The mixing between a $c\bar s$ and a four-quark configuration has been invoked as a mechanism for lowering the mass of $D_{s0}^*(2317)$ \cite{Browder}. Since there is no evidence and strong argument for the existence of tetraquark charmed mesons, it is more plausible to replace the tetraquark configuration by a two-meson state. Indeed, it was first proposed in \cite{vanBeveren} that the low mass of $D_{s0}^*(2317)$ ($D_0^*(2400)$) arises from the mixing between the $0^+$ $c\bar s$ ($c\bar q$) state and the $DK$ threshold ($D\pi$ state). This conjecture has been realized in QCD sum rule and lattice calculations.
It has been shown that when the contribution from the $DK$ continuum is included in QCD sum rules, this effect will significantly lower the mass of the $D_{s0}^*$ state \cite{Dai} and likewise for $D_0^*$ \cite{Dai:D0}. Recent lattice calculations using $c\bar s$, $DK$ and $D^*K$ interpolating fields show the existence of $D_{s0}^*(2317)$ below the $DK$ threshold \cite{Mohler:Ds0} and $D'_{s1}(2460)$ below the $D^*K$ threshold \cite{Lang}. Likewise, a similar lattice calculation with $c\bar q$ and $D\pi$ interpolating fields implies the existence of $D_0^*(2400)$ above the $D\pi$ threshold \cite{Mohler:D0}. All these results indicate that the strong coupling of scalars with hadronic channels will play an essential role of lowering their masses.

In the same spirit, mass shifts of charmed and bottom scalar mesons due to self-energy hadronic loops have been calculated in \cite{Guo}. The results imply that the bare masses of scalar mesons calculated in the quark model can be reduced significantly. Mass shifts due to hadronic loops or strong coupled channels have also been studied in different frameworks to explain the small mass of $D_{s0}^*(2317)$ \cite{Hwang:2004cd,Zhou} and the small mass gap between $D_{s0}^*(2317)$ and its non-strange partner \cite{TLee}.

Since $D_{s0}^*(2317)$ is barely below $DK$ threshold,
it has been pointed out in \cite{Guo:dynamic} that a $DK$ bound state can be dynamically generated in the framework of unitarized chiral perturbation theory with a mass $2312\pm41$ MeV (see also \cite{MartinezTorres}). This $DK$ bound state can be identified with $D_{s0}^*(2317)$. However, there are two $D_0^*$ resonances in this approach: one broad state with a mass $2097\pm18$ MeV and one narrow state with a mass $2448\pm30$ MeV \cite{Guo:dynamic}. Therefore, the non-strange scalar $D_0^*(2400)$ and its closeness to $D_{s0}^*(2317)$ in mass cannot be accounted for in this approach.

The work of Guo, Krewald and Mei\ss ner (GKM) \cite{Guo} is attractive and deserves special attention as they have shown that self-energy contributions can pull down the scalar meson masses significantly. Hence, near degeneracy will be accounted for if the mass shift in the strange sector is larger than that in its non-strange partner by an amount of $80-100$ MeV. GKM considered three different models for calculations. Models I and III correspond to non-derivative and derivative couplings of the scalar meson with two pseudoscalar mesons, while Model II is based on heavy meson chiral perturbation theory (HMChPT).
In this work we will re-examine this mass-shift mechanism. Presumably, calculations based on HMChPT should be less model dependent and thus more reliable. However, we find HMChPT results rather different from GKM. For example, mass shifts in the strange charm sector are found to be largely overestimated. It turns out that the conventional model works better toward the understanding of near mass degeneracy. The scalar $B$ mesons $B_{s0}^*$ and $B_0^*$ have not been observed thus far. In this work we shall examine if near degeneracy observed in the charm sector will imply the similarity of $B_{s0}^*$ and $B_0^*$ masses in the $B$ system, which will be tested experimentally.

This work is organized as follows. In Sec. II we outline the general framework of the mass shift and decay width due to self-energy contributions in the conventional field theory and also in HMChPT. Then we proceed in Sec. III to consider the self-energy corrections to $B_{s0}^*$ as an example in HMChPT. We also calculate mass shifts in the conventional model. The masses of $B_{s0}^*$ and $B_{0}^*$ are discussed in Sec. IV with focus on the predictions based on heavy quark symmetry and possible $1/m_Q$ and QCD corrections. Near degeneracy between $K_0^*(1430)$ and $a_0(1450)$ is discussed in Sec. V. Sec. VI comes to our conclusions.

\section{General framework}
Consider the exact propagator of a scalar meson  \cite{Peskin}
\be
{i\over p^2-m_0^2}+ {i\over p^2-m_0^2}\Big(-i\Pi(p^2)\Big){i\over p^2-m_0^2}+\cdots={i\over p^2-m_0^2-\Pi(p^2)}
\en
with $\Pi(p^2)$ being the 1PI self-energy contribution.
If the scalar particle is a resonance which can decay into two or more light particles, $\Pi(p^2)$ will have an imaginary part.
Therefore, the self-energy $\Pi(p^2)$ is in general complex.
The particle's mass $m$ is then given by the condition
\be \label{eq:phymass}
m^2-m_0^2-{\rm Re}\Pi(m^2)=0.
\en
Writing
\be
\Pi(p^2)=\delta m^2-(Z^{-1}-1)(p^2-m_0^2)+i{\rm Im}\Pi(p^2),
\en
with $Z$ being the wave-function renormalization constant,
we then have
\be
{i\over p^2-m_0^2-\Pi(p^2)}={iZ\over p^2-m_0^2-Z\delta m^2-iZ{\rm Im}\Pi(p^2)}={iZ\over p^2-m^2-iZ{\rm Im}\Pi(p^2)},
\en
where use of Eq. (\ref{eq:phymass}) has been made. In general, the wave-function renormalization constant follows from the prescription
\be
Z=1+\left.{\partial \Pi(s)\over \partial s}\right|_{s=m^2}.
\en
If the resonance is narrow, we can approximate ${\rm Im}\Pi(p^2)$ as ${\rm Im}\Pi(m^2)$ over the width of the resonance. The scalar meson propagator has the Breit-Wigner form, namely,
\be
{i\over p^2-m^2+im\Gamma}.
\en
In this case one can identify the width as
\be
\Gamma=-{Z\over m}{\rm Im}\Pi(m^2).
\en

In this work, chiral loop corrections to the heavy meson's propagator will be also evaluated in the framework of HMChPT (heavy meson chiral perturbation theory) to be described below. The heavy meson's propagator in this framework has the expression
\be
{i\over 2v\cdot k-\Pi(v\cdot k)},
\en
where $v$ and $k$, respectively, are the velocity and  the residual momentum of the meson defined by $p=vm_0+k$.
The particle's on-shell condition is then given by
\be \label{eq:onshell}
2v\cdot \tilde k-{\rm Re}\Pi(v\cdot \tilde k)=0,
\en
which is the analog of Eq. (\ref{eq:phymass}). The physical mass reads
\be
m=m_0+v\cdot \tilde k\,.
\en
Writing
\be
\Pi(v\cdot k)=\delta m-2(Z^{-1}-1)v\cdot k+i{\rm Im}\Pi(v\cdot k),
\en
we are led to
\be
{i\over 2v\cdot k-\Pi(v\cdot k)}={iZ\over 2v\cdot k-Z\delta m-iZ{\rm Im}\Pi(v\cdot k)}={iZ\over 2v\cdot k-2v\cdot \tilde k-iZ{\rm Im}\Pi(v\cdot k)}.
\en
The wave-function renormalization constant $Z$ in general follows from the prescription
\be
Z=1+{1\over 2}\left.{\partial \Pi(v\cdot k)\over \partial v\cdot k}\right|_{v\cdot k\to v\cdot\tilde k}.
\en
The width of a narrow resonance is
\be
\Gamma=-Z\,{\rm Im}\Pi(v\cdot \tilde k).
\en

\section{Mass shift of scalar charmed and bottom mesons due to hadronic loops}

Since chiral loop corrections to heavy scalar mesons have both finite and divergent parts, it is natural to consider the framework of HMChPT where the divergences and the renormalization scale dependence arising from the chiral loops induced by the lowest-order tree Lagrangian can be absorbed into the counterterms which have the same structure as the next-order tree Lagrangian.

\subsection{Hadronic loop in HMChPT}
\begin{figure}[t]
\begin{center}
\includegraphics[width=0.60\textwidth]{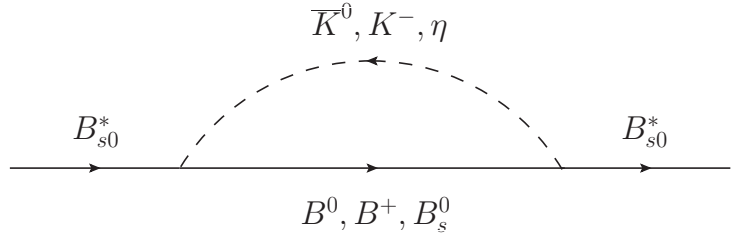}
\vspace{0.0cm}
\caption{Self-energy contributions to the $B_{s0}^*$.} \label{fig:Scalar} \end{center}
\end{figure}

Consider the self-energy contributions of Fig. \ref{fig:Scalar} for the scalar $B_{s0}^*$ meson with $B^0\ov K^0$, $B^+K^-$ and $B_s\eta$ intermediate states. We will evaluate this loop diagram in the framework of HMChPT in which the low energy dynamics of hadrons is described by the formalism in which heavy quark symmetry and chiral symmetry are synthesized \cite{Yan,Wise,Donoghue}. The relevant Lagrangian is (see, e.g. \cite{Casalbuoni})
\be \label{eq:Lagrangian}
{\cal L} &=& {\rm Tr}\left[ \bar H_b(iv\cdot D)_{ba}H_a\right]+
{\rm Tr}\left[\bar S_b((iv\cdot D)_{ba}-\delta_{ba}\Delta_S)S_a\right] \non \\
&&
+h{\rm Tr}\left[\bar S_b\gamma_\mu\gamma_5 {\cal A}^\mu_{ba}H_a\right]+h.c.,
\en
where $H$ denotes the negative-parity spin doublet $(P,P^*)$ and $S$ the positive-parity spin doublet $(P_0,P'_1)$ with $j=1/2$ ($j$ being the total angular momentum of the light degrees of freedom):
\be \label{eq:field}
H_a={ 1+v\!\!\!/\over 2}[P^*_{a\mu}\gamma^\mu-P_a\gamma_5], \qquad
S_a={1\over 2}(1+v\!\!\!/)[{P'}_{1a}^{\mu}\gamma_\mu\gamma_5-P^*_{0a}].
\en
For the self-energy diagram depicted in Fig. \ref{fig:Scalar}, $P_a=(B^+,B^0,B_s^0)$, $P^*_a=(B^{*+},B^{*0},B_s^{*0})$, $P^*_{0a}=(B_0^{*+}, B_0^{*0}, B_{s0}^{*0})$, and
$P'_{1a}=({B'}_1^+,{B'}_1^0,{B'}_{s1}^0)$.  Note that we use the particle symbols to denote the heavy meson field operators $P,P^*,P_0^*,P'_1$, but keep in mind that they differ from the conventional meson fields by a normalization factor of $\sqrt{M}$.
The nonlinear chiral symmetry is realized by making use of the unitary matrix $\Sigma={\rm exp}(i\sqrt{2}\phi/f_\pi)$ with $f_\pi=93$ MeV and $\phi$ being a $3\times 3$ matrix for the octet of Goldstone bosons
\be
\phi\equiv\frac{1}{\sqrt2}\pi^a\lambda^a=\left( \matrix{ {\pi^0\over\sqrt{2}}+{\eta\over\sqrt{6}} & \pi^+ & K^+
   \cr   \pi^- & -{\pi^0\over\sqrt{2}}+{\eta\over\sqrt{6}} & K^0 \cr
                    K^- & \ov K^0 & -\sqrt{2\over 3}\eta  \cr
                  } \right).
                  \en
In terms of the new matrix $\xi=\Sigma^{1/2}$, the axial vector field ${\cal A}$ in Eq. (\ref{eq:Lagrangian}) has the expression
\be
{\cal A}={i\over 2}(\xi^\dagger\partial_\mu\xi-\xi\partial_\mu \xi^\dagger)=-{1\over \sqrt{2}f_\pi}\partial_\mu\phi+{1\over
12\sqrt{2}f_\pi^3}[\phi,[\phi,\partial_\mu\phi]]+\cdots.
\en
The coupling of $B_{s0}^*$ to the $B^0\ov K^0,B^+K^-,B_s\eta$ channels can be read from the interaction term
\be
{\sqrt{2}h\over f_\pi}(P^\dagger_{0b}v^\mu\partial_\mu \phi_{ba}P_a-P^{\nu\dagger}_{1b}v^\mu\partial_\mu \phi_{ba}P_{a\nu}^*).
\en
The vertex of the  $B_{s0}^*B^0\ov K^0$ coupling, for example, is given by
\be \label{eq:vertex}
 -{\sqrt{2}h\over f_\pi}\,(v\cdot q),
\en
where $v$ is the velocity of $B_{s0}^*$ or $B^0$ and $q$ is the momentum of the pseudoscalar meson.
Since in HMChPT, the calculation is done in terms of the field operators $P,P^*,P_0^*,P'_1$, a normalization factor of $\sqrt{M_BM_{B_{s0}^*}}$ should not be incorporated in Eq. (\ref{eq:vertex}), contrary to the Feynman rule for the vertex $P_{0b}^*P_a\phi_{ba}$ given in, for example, \cite{Casalbuoni}.
In Eq. (\ref{eq:Lagrangian}), the parameter $\Delta_S$ is the residual mass of the $S$ field; it measures the mass splitting between positive
and negative parity doublets and can be expressed in terms of the spin-averaged masses
\be
\la M_H\ra\equiv {3M_{P^*}+M_{P}\over 4},\qquad \la M_S\ra\equiv {3M_{P'_1}+M_{P^*_0}\over 4},
\en
so that
\be \label{eq:DeltaS}
\Delta_{S}= \la M_S\ra-\la M_H\ra.
\en

There exist two corrections to the chiral Lagrangian: one from $1/m_Q$ corrections and the other from chiral symmetry breaking.
The $1/m_Q$ corrections are \cite{Casalbuoni,Boyd}
\be
{\cal L}_{1/m_Q}={1\over 2 m_Q}\left\{ \lambda_2^H{\rm Tr}[\bar H_a\sigma^{\mu\nu}H_a\sigma_{\mu\nu}]-\lambda_2^S {\rm Tr}[\bar S_a\sigma^{\mu\nu}S_a\sigma_{\mu\nu}]\right\},
\en
with
\be \label{eq:lambda2}
\lambda_2^H={1\over 4}(M^2_{P^*}-M^2_{P}), \qquad \lambda_2^S={1\over 4}(M^2_{P'_1}-M^2_{P^*_0}),
\en
where $\lambda_H$ ($\lambda_S$) is the mass splitting between spin partners, namely, $P^*$ and $P$ ($P'_1$ and $P^*_0$), of the pseudoscalar (scalar) doublet. We will not write down the explicit expressions for chiral symmetry breaking terms and the interested reader is referred to \cite{Mehen}.
The masses of heavy mesons can be expressed as
\be \label{eq:massrel}
M_{P_a} &=& M_0-{3\over 2}{\lambda_2^H\over m_Q}+\Delta_a, \qquad\qquad M_{P^*_a} = M_0+{1\over 2}{\lambda_2^H\over m_Q}+\Delta_a, \non \\
M_{P_{0a}^*} &=& M_0+\Delta_S-{3\over 2}{\lambda_2^S\over m_Q}+\tilde\Delta_a, \quad~ M_{P'_{1a}} = M_0+\Delta_S+{1\over 2}{\lambda_2^S\over m_Q}+\tilde\Delta_a,
\en
where we have followed \cite{Fajfer} to use $\Delta_a$ and $\tilde\Delta_a$ to denote the residual mass contributions to negative- and
positive-parity mesons, respectively. In heavy quark effective theory, the parameter $M_0$ in Eq. (\ref{eq:massrel}) is related to $m_Q+\bar\Lambda _{H_Q}$, see Eq. (\ref{eq:HQET}).
Note that $\lambda_2^H/m_Q\approx {1\over 2}(M_{P^*}-M_P)\equiv {1\over 2}\Delta M_P$ and $\lambda_2^S/m_Q\approx {1\over 2}\Delta M_S$ in the heavy quark limit. The propagators for $P_a(v)$ and $P_{0a}^*(v)$ read
\be
{i\over 2(v\cdot k+{3\over 4}\Delta M_P-\Delta_a)+i\epsilon}, \qquad {i\over 2(v\cdot k-\Delta_S +{3\over 4}\Delta M_S-\tilde\Delta_{a})+i\epsilon},
\en
respectively.

Consider the hadronic loop contribution to $B_{s0}^*$ in Fig. \ref{fig:Scalar} with the intermediate states $B^0$ and $\ov K^0$.
The self-energy loop integral is
\be \label{eq:loop}
\Pi(v\cdot k) &=& \left({2h^2\over f_\pi^2}\right){i\over 2}\int {d^4q\over (2\pi)^4}\,{(v\cdot q)^2\over (q^2-m^2+i\epsilon)(v\cdot k'+{3\over 4}\Delta M_B-\Delta_u+ i\epsilon)} \non \\
&=& \left({2h^2\over f_\pi^2}\right){i\over 2}\int {d^4q\over (2\pi)^4}\,{(v\cdot q)^2\over (q^2-m^2+i\epsilon)(v\cdot q+\omega+i\epsilon)},
\en
where $m$ is the mass of the Goldstone boson.
The residual momentum $k'$ of the heavy meson in the loop is given by
$k'=p+q-vM_B=q+k+v({\cal M}_{B_{s0}^*}-M_B)$,
and $\omega=v\cdot k+{\cal M}_{B_{s0}^*}-M_B+{3\over 4}\Delta M_B-\Delta_u$. We shall use the calligraphic symbol  ${\cal M}_{B_{s0}^*}$ to denote the bare mass of $B_{s0}^*$.
The full propagator becomes (the $i\epsilon$ term being dropped for convenience)
\be
{i\over 2(v\cdot k-\Delta_{S}+{3\over 4}\Delta M_S-\tilde\Delta_s) -[2\,\Pi_{BK}(v\cdot k)+{2\over 3}\,\Pi_{B_s\eta}(v\cdot k)]}
\en
for $B_{s0}^*$ after taking into account the contributions from the channels $B^0\ov K^0,B^+K^-,B_s^0\eta$  and likewise,
\be
{i\over 2(v\cdot k-\Delta_{S}+{3\over 4}\Delta M_S-\tilde\Delta_u) -[{3\over 2}\,\Pi_{B\pi}(v\cdot k)+{1\over 6}\,\Pi_{B\eta}(v\cdot k)+\Pi_{B_sK}(v\cdot k)]}
\en
for $B_0^{*+}$. Since the parameters $\Delta_S$, $\Delta M_S$ and $\tilde\Delta_s$ are unknown, we are not able to determine mass shifts from above equations. Assuming that the bare mass ${\cal M}$ is the one obtained in the quark model, then from Eq. (\ref{eq:massrel}) we have
\be
\Delta_{S}-{3\over 4}\Delta M_S+\tilde\Delta_s={\cal M}_{B_{s0}^*}-M_0={\cal M}_{B_{s0}^*}-M_B-{3\over 4}\Delta M_B+\Delta_u, \non \\
\Delta_{S}-{3\over 4}\Delta M_S+\tilde\Delta_u={\cal M}_{B_{0}^*}-M_0={\cal M}_{B_{0}^*}-M_B-{3\over 4}\Delta M_B+\Delta_u,
\en
for $B_{s0}^*$ and $B_0^*$, respectively.
With
\be
&& F_1(v\cdot k)\equiv 2(v\cdot k-{\cal M}_{B_{s0}^*}+M_B+{3\over 4}\Delta M_B-\Delta_u)-{\rm Re}\left[2\,\Pi_{BK}(v\cdot k)+{2\over 3}\,\Pi_{B_s\eta}(v\cdot  k)\right],  \\
&& F_2(v\cdot k)\equiv 2(v\cdot k-{\cal M}_{B_{0}^*}+M_B+{3\over 4}\Delta M_B-\Delta_u)-{\rm Re}\left[{3\over 2}\,\Pi_{B\pi}(v\cdot k)+{1\over 6}\,\Pi_{B\eta}(v\cdot  k)+\Pi_{B_sK}(v\cdot  k)\right], \non
\en
the on-shell conditions read
\be \label{eq:vdotk}
F_1(v\cdot \tilde k)=0, \qquad F_2(v\cdot \tilde k)=0.
\en
Obviously, $v\cdot \tilde k$ is different for $B_{s0}^*$ and $B_0^*$.
The physical mass is then given by
\be
M_{B_{(s)0}^*}=M_0+v\cdot \tilde k=M_B+{3\over 4}\Delta M_B-\Delta_u+v\cdot \tilde k\,,
\en
while the width of the $B_0^*$ reads
\be
\Gamma=-{3\over 2}Z\,{\rm Im}\Pi_{B\pi}(v\cdot\tilde k).
\en
Since $\Delta_u$ is of order 1 MeV \cite{Stewart}, it can be neglected in practical calculations.

The loop integral in Eq. (\ref{eq:loop}) has been evaluated in the literature from time to time \cite{Cho,Cheng:SU(3),Falk,Boyd,Bernard,Stewart,Scherer}. The complete result is \cite{Falk,Boyd,Stewart}
\be \label{eq:G}
\Pi(v\cdot k)= \left({2h^2\over f_\pi^2}\right){\omega\over 32\pi^2}\left[ (m^2-2\omega^2){\rm ln}{\Lambda^2\over m^2}-2\omega^2+4\omega^2 F\left(-{m\over \omega}\right)\right],
\en
with
\be \label{eq:F}
F\left({1\over x}\right)=\cases{ {\sqrt{x^2-1}\over x}\,{\rm ln}(x+\sqrt{x^2-1}), & $|x|\geq 1$ \cr
-{\sqrt{1-x^2}\over x}\left[{\pi\over 2}-{\rm tan}^{-1}\left({x\over\sqrt{1-x^2}}\right)\right], & $|x|\leq 1$\,. }
\en
As emphasized in \cite{Stewart}, the expression of $F(x)$ in \cite{Falk,Boyd} should be replaced by Eq. (\ref{eq:F}). The logarithm in the above equation has an imaginary part  when $x=-\omega/m<1$, or when $v\cdot\tilde k+{\cal M}_{B_{s0}^*}=M_{B_{s0}^*}>M_P+m$, i.e. the imaginary part occurs when the intermediate states can be put on shell.

Note that a different expression for the loop integral in $\Pi(v\cdot k)$ was obtained in \cite{Scherer}
\be \label{eq:otherG}
\Pi(v\cdot k)=\left({2h^2\over f_\pi^2}\right){m^2\over 16\pi^2}J(\omega),
\en
with
\be
J(\omega)=-\omega\left({\rm ln}{\Lambda^2\over m^2}+1\right)+\cases{ 2\sqrt{\omega^2-m^2}\,{\rm cosh}^{-1}({\omega\over m})-2\pi i\sqrt{\omega^2-m^2}, & $\omega>m$ \cr
2\sqrt{m^2-\omega^2}\,{\rm cos}^{-1}(-{\omega\over m}), & $\omega^2<m^2$ \cr
-2\sqrt{\omega^2-m^2}\,{\rm cosh}^{-1}(-{\omega\over m}), & $\omega < -m$\,. }
\en
This expression disagrees with Eq. (\ref{eq:G}), for example, the coefficient of the term $m^2\omega\, {\rm ln}{\Lambda^2\over m^2}$ differs in sign and magnitude from Eq. (\ref{eq:G}). A crucial difference between the two expressions of $\Pi(v\cdot k)$ given by  Eqs. (\ref{eq:otherG}) and (\ref{eq:G}) is that the loop integral of the former vanishes in the chiral limit, while it is not the case for the latter.

\begin{figure}[t]
\centering
  \includegraphics[width=0.43\textwidth]{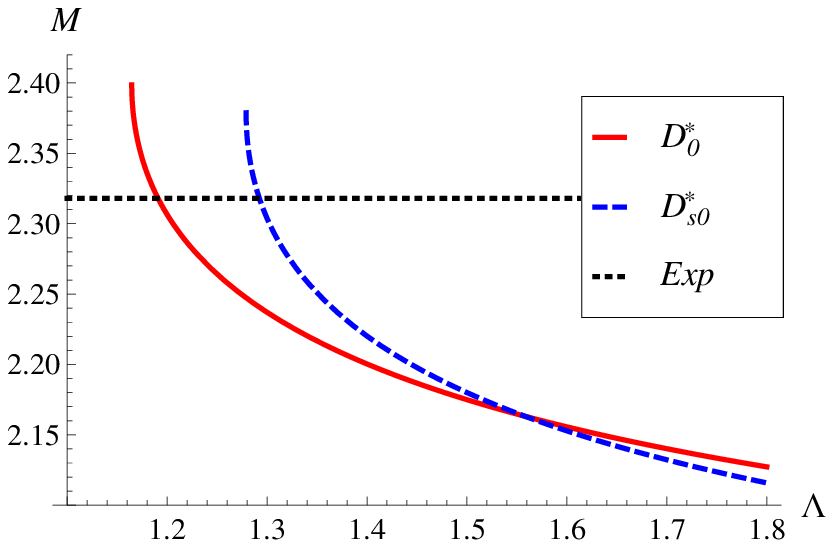}
 \hspace{0.2in}
  \includegraphics[width=0.43\textwidth]{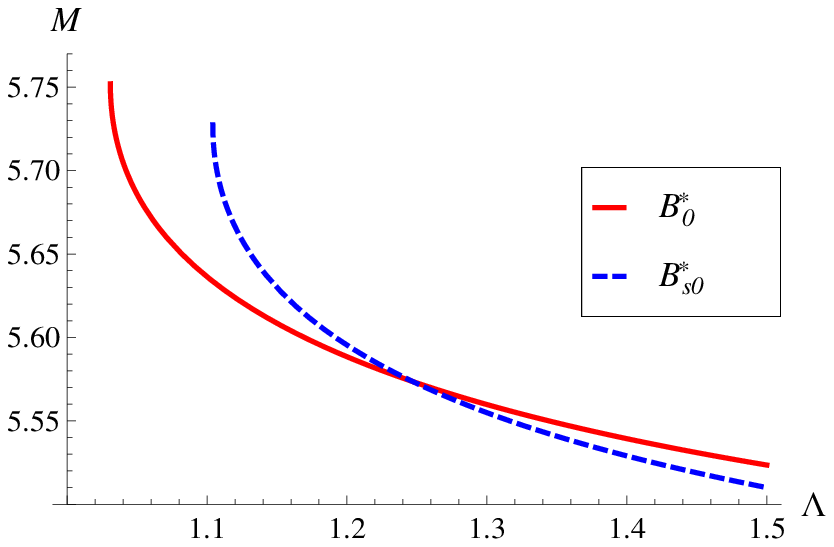}
\caption{Physical masses of $D_{s0}^*$ and $D_0^*$ (left panel) as well as $B_{s0}^*$ and $B_0^*$ (right panel) calculated in HMChPT as a function of the renormalization scale $\Lambda$. Bare masses are taken from \cite{Godfrey}.
All the scales are in units of GeV.}
\label{fig:MvsLambda}
\end{figure}

The parameter $\Lambda$ appearing in Eq. (\ref{eq:G}) is an arbitrary renormalization scale. In the dimensional regularization approach, the common factor
\be
{2\over \epsilon}-\gamma_E+{\rm ln}4\pi+1
\en
with $\epsilon=4-n$ can be lumped into the logarithmic term ${\rm ln}(\Lambda^2/m^2)$. Of course, physical amplitudes should be independent of the renormalization scale. This means that the $\Lambda$ dependence from the chiral loop must be exactly compensated by the $\Lambda$ dependence of local counterterms in higher order chiral Lagrangian. In the conventional practice, it is often to choose $\Lambda\sim \Lambda_\chi$, the chiral symmetry breaking scale of order 1 GeV, to get numerical estimates of chiral loop effects. However, we find that for $\Lambda\sim 1$ GeV, the two equations in (\ref{eq:vdotk}) do not have solutions. This can be seen by plotting $F_1(v\cdot k)$ and $F_2(v\cdot k)$ as a function of $v\cdot k$. It turns out that neither curve in a parabola shape that opens downward has a $v\cdot k$-intercept. Both curves are shifted up
with increasing $\Lambda$ and will eventually intercept with the $v\cdot k$ axis at a critical renormalization scale.  Fig. \ref{fig:MvsLambda} shows the dependence of the physical masses of heavy scalar mesons on the renormalization scale $\Lambda$. It is clear that the critical renormalization scale is, for example, 1.17 GeV for $D_{0}^*$ and 1.30 GeV for $D_{s0}^*$. In general, there exist two solutions for $v\cdot\tilde k$ due to two intercepts of the curve with the $v\cdot k$ axis. We shall consider the smaller solution for $v\cdot\tilde k$ as the other solution will yield too large masses.

\begin{table}[t]
\caption{Mass shifts ($\delta M\equiv M-{\cal M}$) and widths ($\Gamma$) of heavy scalar mesons calculated in HMChPT. The renormalization scale is taken to be $\Lambda=1.3$ (1.2) GeV for scalar $D$ ($B$) mesons. Bare masses are taken from \cite{Godfrey,Di Pierro}. All masses and widths are given in MeV and only the central values are listed here.} \label{tab:prediction}
\begin{ruledtabular}
\begin{tabular}{|c | c  c c c | c c c c| }
Meson  &   Bare mass \cite{Di Pierro} & $\delta M$ & $M$ & $\Gamma$~~ & Bare mass \cite{Godfrey} & $\delta M$ & $M$ & $\Gamma$~~~  \\ \hline
$B_{s0}^*$ &   5804  & $-236$ &  5568 & & 5830  & $-235$ & 5595 &   \\
$B_{0}^*$ &  5706  &  $-158$ & 5548 &  55~~~ & 5760 & $-172$ & 5588 &  110~~~  \\
$D_{s0}^*$ & 2487  & $-146$ &  2341 & & 2480  & $-176$ & 2304 &  \\
$D_0^*$ &  2377  & $-165$ &  2212 & 130~~~ & 2400  & $-163$ & 2237 & 183~~~  \\
\end{tabular}
\end{ruledtabular}
\end{table}

For numerical calculations, we follow \cite{Colangelo} to use $h=0.56\pm0.04$, which is the weighted average of the coupling constant $h$ extracted from the measured widths of $D_0^{*0},D_0^{*\pm}$ and ${D'}_1^{0}$.
The coupled channels are $D^0K^+,D^+K^0,D_s^+\eta$ for $D_{s0}^*$ and $D^+\pi^0,D^0\pi^+,D^+\eta,D_s^+K^0$ for $D_0^{*+}$.
The results are displayed in Table \ref{tab:prediction} for some specific values of $\Lambda$. \footnote{
Although our study of mass shift effects due to chiral loops in HMChPT is very
similar to GKM \cite{Guo}, our calculations differ form theirs in
several aspects: (i)  We have applied Eq. (\ref{eq:G}) rather than (\ref{eq:otherG}) to evaluate the loop integral in Eq. (\ref{eq:loop}). It was claimed by GKM that the mass shift vanishes in the chiral limit, while it is not the case in our analysis. (ii) The additional normalization factor of $\sqrt{M_D \buildrel {\rm o}\over {M}_{D_0^*}}$ in Eq. (12) of \cite{Guo} for the $D_0^*D\pi$  coupling should not be included in Eqs. (17)-(19) for the scalar meson propagators as the calculation is done in terms of the field operators $P$ and $P_0^*$. Indeed, one can check that the second term in their Eq. (18) does not have the correct dimension.
(iii) The propagator of the scalar meson should include the mass splitting $\Delta_S$ defined in Eq. (\ref{eq:DeltaS}). The term $2(v\cdot p-\buildrel {\rm o}\over {M}_{D_{s0}^*})$  in Eq. (18) of \cite{Guo} is not equivalent to $(2v\cdot k+{3\over 4}\Delta_{S_s}-\delta_S)$ in Eq. (17) of the same reference.}
From Fig. \ref{fig:MvsLambda} it is evident that near mass degeneracy occurs when $\Lambda$ is in the vicinity of 1.56 (1.25) GeV for the scalar $D$ ($B$) sector. However, the corresponding mass of order 2160 MeV for scalar $D$ mesons is too small compared to the experimental value of 2317 MeV. By contrast, the measured mass of $D_{s0}^*$ can be reproduced by having $\Lambda$ around 1.3 GeV, but the predicted $D_0^*$ will be too light.
Evidently, self-energy corrections in general will push down the masses of $B_{s0}^*$ much more than that of $B_0^*$, but this is not the case for $D_{s0}^*$ and $D_0^*$.
It is not clear why HMChPT does not lead to satisfactory results and this is probably ascribed to the leading heavy quark expansion we adapted here. It could be that higher-order heavy quark expansion needs to be taken into account to justify the use of $\Lambda\sim \Lambda_\chi$ (see also the discussion in \cite{Guo}).

\subsection{Mass shift in the conventional model}

Since HMChPT does not lead to satisfactory results, we shall perform the calculation in the conventional model without appealing to the heavy quark expansion; that is, it is based on full theory rather than effective theory. In general, the coupling of a scalar meson to two light pseudoscalar mesons can be of derivative or non-derivative type:
\be \label{eq:SPPcoupling}
{\cal L}_1=f_{SPP}S^i_j P^j_k P^k_i, \qquad\quad {\cal L}_2=g_{SPP}S^i_j\partial_\mu P^j_k\partial^\mu P^k_i,
\en
where $S$ and $P$ denote the nonet states of scalar and pseudoscalar mesons, respectively. However, the derivative-type coupling is preferred as it follows from a chiral invariant Lagrangian in which $P^i_j$ transforms non-linearly under axial transformations \cite{Black}. Moreover, it is also the form implied by HMChPT. We shall write the vertex for the $B_{s0}^*\ov K^0(q)B^0(q')$ as
\be
-{\sqrt{2}g\over f_\pi}\,(q\cdot q').
\en
Comparing this with Eq. (\ref{eq:vertex}) and using the fact that the heavy meson operator in HMChPT differs from the conventional heavy meson field by a normalization factor of $\sqrt{M}$, we see that the unknown coupling $g$ should be similar to the coupling $h$ and is independent of the heavy flavor.

The relevant chiral loop is
\be
\Pi(s) &=& i\left({2g^2\over f_\pi^2}\right)\int {d^4q\over (2\pi)^4}{[q\cdot (p+q)]^2\over (q^2-m_1^2+i\epsilon)[(p+q)^2-m_2^2+i\epsilon]}
\en
with $s=p^2$. This loop integral has the expression \cite{Guo} \footnote{It is necessary to separate the logarithmic terms to get the imaginary part, though it is fine to combine the logarithmic terms to obtain the real part of $\Pi(s)$.}
\be \label{eq:loops}
\Pi(s)&=& \left({2g^2\over f_\pi^2}\right){1\over 16\pi^2}\Bigg\{-{\rm ln}{\Lambda^2\over m_2^2}\left[m_1^4+m_1^2m_2^2+m_2^4-{3\over 4}(m_1^2+m_2^2)s+{s^2\over 4}\right] \non \\
&& -{(m_1^2+m_2^2-s)^2\over 4}
 +{(m_1^2-m_2^2+s)[\lambda+2m_1^2(m_1^2+m_2^2)]-2m_1^2\lambda\over 8s}\,{\rm ln}{m_1^2\over m_2^2}
 \non \\
&& +{\sqrt{\lambda}(m_1^2+m_2^2-s)^2\over 8s}\Big[{\rm ln}(s-m_1^2+m_2^2+\sqrt{\lambda}) -{\rm ln}(-s+m_1^2-m_2^2+\sqrt{\lambda})
\non \\
&& +{\rm ln}(s+m_1^2-m_2^2+\sqrt{\lambda})-{\rm ln}(-s-m_1^2+m_2^2+\sqrt{\lambda})\Big]\Bigg\},
\en
where $\lambda=[s-(m_1+m_2)^2][s-(m_1-m_2)^2]$ is the so-called K\"all\'en function.
The physical masses of $B_{s0}^*$ and $B_0^*$ are determined  by the equations (see Eq. (\ref{eq:phymass}))
\be
&& M^2_{B_{s0}^*}-{\cal M}^2_{B_{s0}^*}-{\rm Re}\,\left[2\,\Pi_{BK}(M^2_{B_{s0}^*})+{2\over 3}\,\Pi_{B_s\eta}(M^2_{B_{s0}^*}) \right]=0, \non \\
&& M^2_{B_{0}^*}-{\cal M}^2_{B_{0}^*}-{\rm Re}\,\left[{3\over 2}\,\Pi_{B\pi}(M^2_{B_{0}^*})+{1\over 6}\,\Pi_{B\eta}(M^2_{B_{0}^*})+\Pi_{B_sK}(M^2_{B_{0}^*})\right]=0,
\en
while the width of $B_0^*$ arising from the imaginary part of the $B\pi$ loop reads
\be \label{eq:B0width}
\Gamma=-{3\over 2}{1\over M_{B_{0}^*}}Z\,{\rm Im}\,\Pi_{B\pi}(M_{B_{0}^*}^2).
\en

\begin{table}[t]
\caption{Same as Table \ref{tab:prediction} except in the conventional model, where the renormalization scale $\Lambda$ is set to be 1.85 GeV in the $D$ system and $5.28$ GeV in the $B$ sector.  Note that the masses of $D_{s0}^*(2317)$ and $D_0^*(2400)$ can be well reproduced by choosing the $\Lambda$ scale slightly different for $D_{s0}^*$ and $D_0^*$; see the main text.
} \label{tab:fieldcal}
\begin{ruledtabular}
\begin{tabular}{|c | c  c c c | c c c c| }
Meson  &   Bare mass \cite{Di Pierro} & $\delta M$ & $M$ & $\Gamma$~~~ & Bare
 mass \cite{Godfrey} & $\delta M$ & $M$ & $\Gamma$~~~  \\ \hline
$B_{s0}^*$ &   5804  & $-85$ &  5719& & 5830  & $-105$ & 5725 &   \\
$B_{0}^*$ &  5706  &  $4$ & 5710 &  68~~~ & 5760 & $-37$ & 5723 &  71~~~  \\
$D_{s0}^*$ & 2487  & $-130$ &  2357 & & 2480  & $-124$ & 2356 &  \\
$D_0^*$ &  2377  & $-72$ &  2305 & 124~~~ & 2400  & $-83$ & 2317 & 131~~~  \\
\end{tabular}
\end{ruledtabular}
\end{table}

The strong coupling constant $g$ is determined from the measured
partial width through the relation
 \be
 \Gamma_{S\to P_1P_2}={p_c\over 16\pi f_\pi^2M_S^2}g_{S\to P_1P_2}^2(M_S^2-m_1^2-m_2^2)^2,
 \en
with $p_c$ being the c.m. momentum. We obtain
\be
 g=0.55\pm0.04
\en
from the experimental measurement $\Gamma(D_0^*(2400)^0)=267\pm40\,{\rm MeV}$ (see Table \ref{tab:data}).
As for the renormalization scale $\Lambda$, it was assumed in \cite{Guo} that at some subtraction point the contribution from any hadronic loop vanishes. We will follow \cite{Guo} to take the substraction scale at $M_D$ and $M_B$ for the charm and bottom sectors, respectively. The corresponding renormalization scales are $\Lambda=1.85$ and 5.28 GeV.

Table \ref{tab:fieldcal} shows the calculational results. \footnote{Note that although our calculation is the same as Model III of GKM \cite{Guo}, the numerical results are somewhat different especially for the $B$ sector. For example, when the bare masses are chosen to be ${\cal M}_{B^*_{s0}}=5830$ MeV and ${\cal M}_{B^*_{0}}=5760$ MeV, we obtain $\delta M=-105$ MeV for $B^*_{s0}$ and $-37$ MeV for $B^*_{0}$ (see Table \ref{tab:fieldcal}), while the corresponding results are $-(130\sim 210)$ MeV and $-(120\sim 210)$ MeV in \cite{Guo}.}
It is obvious that the conventional model works better than HMChPT for mass shifts induced by self energies.
We see that the mass shift in the strange sector is large enough to explain the small mass of $D_{s0}^*$ and the smaller mass gap between $D_{s0}^*$ and $D_0^*$ than naive expectation. We notice that for given bare masses, say, the ones in \cite{Godfrey}, the measured masses of $D_{s0}^*(2317)$ and $D_0^*(2400)$ can be well reproduced by choosing $\Lambda$ to be 1.93 GeV for $D_{s0}^*$ and 1.85 GeV for $D_0^*$.
Moreover, we learn from Table \ref{tab:fieldcal}  that near mass degeneracy should work even better in the $B$ sector.

It should be stressed that chiral loop calculations performed here are meant to demonstrate that the mass difference between strange heavy scalar mesons and their non-strange partners can be substantially reduced by self-energy contributions. The predicted physical masses of scalar mesons are subject to many uncertainties: (i) the unknown input bare masses ${\cal M}$ which are taken from the quark model calculation in this work, \footnote{Conversely, one can use the measured physical masses and chiral loop corrections to deduce the bare masses. For example, it was obtained ${\cal M}_D=1971$ MeV and ${\cal M}_{D^*}=2112$ MeV  in \cite{Cheng:SU(3)} based on HMChPT. Hence, mass shifts of $D$ and $D^*$ are $-107$ and $-102$ MeV, respectively.} (ii) contributions from other intermediate states, for example, the channels $B_0^{*0}\ov K^0$, $B^{*0}\ov K^{*0}$, $B_s\eta',\cdots$ in Fig. \ref{fig:Scalar}.  Presumably, the additional contributions are suppressed as they are more far away from the threshold. (iii) the model-dependent choice of the renormalization scale $\Lambda$.  Of course, physics should be independent of the renormalization scale. However, this issue cannot be properly addressed in the phenomenological model discussed in this section. In view of many theoretical uncertainties, we should not rely on the chiral loop calculations to make quantitative predictions on the scalar meson masses.

\section{Masses of $B^*_{(s)0}$ and $B'_{(s)1}$ mesons}
The states $B^*_{(s)0}$ and $B'_{(s)1}$ have not yet been observed.
The predictions of their masses in the literature are summarized in Table \ref{tab:literature}. Basically, they can be classified into the following categories:
(i) relativistic quark potential model  \cite{Di Pierro,Ebert,Zhu,Godfrey},
(ii) potential model with one loop corrections \cite{Lee,Swanson},
(iii) semi-relativistic quark potential model \cite{Matsuki},
(iv) heavy meson chiral perturbation theory \cite{Bardeen,Colangelo},
(v) chiral quark-pion Lagrangian with strong coupled channels \cite{Badalian},
(vi) unitarized chiral perturbation theory \cite{Rincon,Cleven:2010aw,Guo:dynamic,Cleven:2014},
(vii) QCD sum rules \cite{ZGWang},
(viii) lattice \cite{Koponen},
and (ix) others, such as the MIT bag model \cite{Orsland} and the nonlinear chiral SU(3) model \cite{Lutz:2003fm}.
It is clear that within the framework of the constitute quark model, the mass difference between strange and non-strange scalar $B$ mesons is of order $80-110$ MeV and $B_{s0}^*$ is above $BK$ threshold. Note that the predictions of \cite{Di Pierro} are somewhat smaller than that of \cite{Ebert,Zhu,Godfrey} and that the predicted mass difference between $B_{s0}^*$ and $B_0^*$ can be even larger than 200 MeV in the approach of unitarized chiral perturbation theory \cite{Rincon,Guo:dynamic}. \footnote{This is not a surprise. As mentioned in the Introduction, the non-strange scalar $D_0^*(2400)$ and its closeness to $D_{s0}^*(2317)$ in mass cannot be accounted for by unitarized ChPT; the predicted broad $D_0^*$ state is too light by 220 MeV \cite{Guo:dynamic}.}

As we have argued before, the near mass degeneracy between $B_0^*$ and $B_{s0}^*$ is expected to work even better in the $B$ meson sector. A scrutiny of Table \ref{tab:literature} shows that this degeneracy is respected only in the predictions of \cite{Lee}, \cite{Colangelo} and \cite{Dmitrasinovic:2012zz}.
Since the calculated masses $M_{B_1}=5709$ MeV, $M_{B_2^*}=5723$ MeV, $M_{B_{s1}}=5798$ MeV and $M_{B_{s2}^*}=5813$ MeV in \cite{Lee} are smaller than the data by order 20 and 30 MeV, respectively, this implies that the predictions in \cite{Lee} are probably less reliable than that in \cite{Colangelo} and \cite{Dmitrasinovic:2012zz}. Defining the mass differences between $0^+$ and $0^-$ states and between $1^+$ and $1^-$ states  by
\be
&& \Delta_{s0}^{(c\bar s)}\equiv M_{D_{s0}^*}-M_{D_s}, \qquad \Delta_{0}^{(c\bar q)}\equiv M_{D_{0}^*}-M_{D}, \non \\
&& \Delta_{s1}^{(c\bar s)}\equiv M_{D'_{s1}}-M_{D_s^*}, \qquad \Delta_{1}^{(c\bar q)}\equiv M_{D'_{1}}-M_{D^*},
\en
and {\it assuming} that they are heavy flavor independent, namely,
\be
\Delta_{s0}^{(b\bar s)}=\Delta_{s0}^{(c\bar s)}, \qquad \Delta_{0}^{(b\bar q)}=\Delta_{0}^{(c\bar q)}; \qquad \Delta_{s1}^{(b\bar s)}=\Delta_{s1}^{(c\bar s)}, \qquad \Delta_{1}^{(b\bar q)}=\Delta_{1}^{(c\bar q)},
\en
the author of \cite{Dmitrasinovic:2012zz} obtained the masses of $B^*_{(s)0}$ and $B'_{(s)1}$ as listed in Table \ref{tab:literature}. \footnote{
We have updated the predictions in Table IV of \cite{Dmitrasinovic:2012zz} by using the 2012 PDG \cite{PDG}. The original numbers quoted there are (in MeV) $M_{B_0^*}=5718$, $M_{B_{s0}^*}=5719$,
$M_{B'_1}=5732$, $M_{B'_{s1}}=5765$ with uncertainty of 25 MeV.}

\begin{table}[t]
\centering \caption{Predicted masses (in MeV) of $B^*_{(s)0}$ and $B'_{(s)1}$ mesons in the literature. One broad and one narrow states of $B_0^*$ are found in \cite{Guo:dynamic} and only the mass of the broad state is listed here. The predictions in \cite{Dmitrasinovic:2012zz} have been updated using the 2012 PDG \cite{PDG}; see the main text.} \label{tab:literature}
\begin{tabular}{|c   c c | c c  | }\hline
Ref. \hskip 1 cm  & $B_0^*$ & $B'_1$ & $B_{s0}^*$ & $B'_{s1}$  \\ \hline
\cite{Cleven:2014} & & & $5625\pm45$ & $5671\pm45$ \\ \hline
\cite{Lutz:2003fm} & 5526 &  5590 & 5643 & 5690 \\ \hline
\cite{Rincon} & 5530 & 5579 & 5748 & 5799 \\ \hline
\cite{Guo:dynamic,Guo:Ds1} & $5536\pm29$ & & $5725\pm39$ & $5778\pm7$\\ \hline
\cite{Matsuki} & 5592 & 5649 & 5617 & 5682 \\ \hline
\cite{Orsland} & 5592 & 5671 & 5667 & 5737 \\ \hline
\cite{Vijande} & 5615 & & 5679 & 5713 \\ \hline
\cite{Cleven:2010aw} & & & $5696\pm40$ & $5742\pm40$ \\ \hline
\cite{Bardeen} & $5627\pm35$ & $5674\pm35$ & $5718\pm35$ & $5765\pm35$  \\ \hline
\cite{Lee} & 5637 & 5673 & 5634 & 5672 \\ \hline
\cite{Badalian} & $5675\pm20$ & $5725\pm20$ & $5710\pm15$ & $5730\pm15$ \\
\hline
\cite{Riska} & 5678 & 5686 & 5781 & 5795 \\ \hline
\cite{ZGWang} & & & $5700\pm110$ & $5720\pm90$ \\ \hline
\cite{Di Pierro} & 5706 & 5742 & 5804 & 5842 \\ \hline
\cite{Colangelo} & $5708.2\pm22.5$ \hskip 2 cm & $5753.3\pm31.1$ \hskip 2 cm & ~$5706.6\pm1.2$~ \hskip 2 cm  & $5765.6\pm1.2$ \hskip 2 cm \\ \hline
\cite{Koponen} & & & $5760\pm9$ & $5807\pm9$ \\ \hline
\cite{Dmitrasinovic:2012zz} & $5728\pm25$ & $5742\pm25$ & $5716\pm25$ & $5763\pm25$ \\ \hline
\cite{Swanson} & 5730 & 5752 & 5776 & 5803 \\ \hline
\cite{Ebert} & 5749 & 5774 & 5833 & 5865 \\ \hline
\cite{Zhu} & 5756 & 5779 & 5830 & 5858 \\ \hline
\cite{Godfrey} & 5760 & 5780 & 5830 & 5860 \\ \hline
\end{tabular}
\end{table}


The calculation in  \cite{Colangelo} is based on the heavy quark symmetry argument. In the heavy quark limit, the two parameters $\Delta_S$ and $\lambda_2^S$ defined in Eqs. (\ref{eq:DeltaS}) and (\ref{eq:lambda2}), respectively,
are independent of the heavy quark flavor,
where $\Delta_S$ measures the spin-averaged mass splitting between the scalar doublet $(P'_1,P^*_0)$ and the pseudoscalar doublet $(P^*,P)$ and $\lambda_S$ is the mass splitting between spin partners, namely $P'_1$ and $P^*_0$, of the scalar doublet. From the data listed in Table \ref{tab:data}, we are led to
\be \label{eq:DeltaScq}
\Delta_S^{(c\bar u)}=429\pm28\,{\rm MeV},\qquad \lambda_{2}^{S(c\bar u)}=(359\pm76\,{\rm MeV})^2
\en
for the scalar $c\bar u$ meson and
\be  \label{eq:DeltaScs}
\Delta_S^{(c\bar s)}=347.7\pm0.6\,{\rm MeV},\qquad \lambda_{2}^{S(c\bar s)}=(411.4\pm1.3\,{\rm MeV})^2
\en
for the $c\bar s$ meson. The large errors associated with $\Delta_S$ and $\lambda_2^S$ in the non-strange scalar meson sector reflect the experimental difficulty in identifying the broad states.
Note that
\be
\Delta_S^{(c\bar s)}={1\over 4}\left(3\Delta_1^{(c\bar s)}+\Delta_0^{(c\bar s)}\right), \qquad
\Delta_S^{(c\bar q)}={1\over 4}\left(3\Delta_1^{(c\bar q)}+\Delta_0^{(c\bar q)}\right).
\en
The light quark flavor dependence of $\Delta_S$ and $\lambda_2^S$ shown in Eqs. (\ref{eq:DeltaScq}) and (\ref{eq:DeltaScs})
indicates SU(3) breaking effects. The heavy quark flavor independence of $\Delta_S$ and $\lambda_2^S$ implies
\be \label{eq:HQSrel}
\Delta_S^{(b\bar q)}=\Delta_S^{(c\bar q)},\quad \Delta_S^{(b\bar s)}=\Delta_S^{(c\bar s)}; \qquad
\lambda_{2}^{S(b\bar q)}=\lambda_{2}^{S(c\bar q)}, \quad \lambda_{2}^{S(b\bar s)}=\lambda_{2}^{S(c\bar s)}.
\en
This leads to the predictions of the $B^*_{(s)0}$ and $B'_{(s)1}$ masses (in MeV) \cite{Colangelo}
\be \label{eq:BmassHQS}
&& M_{B_0^*}=5708.2\pm22.5, \qquad M_{B_{s0}^*}=5706.6\pm1.2,  \non \\
&& M_{B'_1}=5753.3\pm31.1, \qquad M_{B'_{s1}}=5765.6\pm1.2 \,.
\en
Evidently, mass degeneracy in the scalar $D$ sector will repeat itself in the $B$ sector through heavy quark symmetry.  Note that in the framework of unitarized HMChPT, $DK$ and $D^*K$ bound states can be dynamically generated with masses in agreement with $D_{s0}^*(2317)$ \cite{Guo:dynamic} and $D'_{s1}(2460)$ \cite{Guo:Ds1}, \footnote{The effect of the nearby $D^*K$ thereshold on $D'_{s1}(2460)$ was recently studied using lattice QCD \cite{Lang}.}
respectively. It is thus interesting to see what are the predictions in this approach for the $B$ analogs. It turns out that the predicted masses $M_{B^*_{s0}}=5725\pm39$ MeV \cite{Guo:dynamic}, $M_{B'_{s1}}=5778\pm7$ MeV \cite{Guo:Ds1}, and $M_{B^*_{s0}}=5696\pm40$ MeV, $M_{B'_{s1}}=5742\pm40$ MeV \cite{Cleven:2010aw} (the last two being obtained in unitarized HMChPT at next-to-leading order) are consistent with the HQS (heavy quark symmetry) results (\ref{eq:BmassHQS}) within errors.

The relations shown in Eq. (\ref{eq:HQSrel}) are valid in the heavy quark limit and they do receive $1/m_Q$ and QCD corrections. We first discuss the corrections to the relation $\lambda_2^{S(b\bar q)}=\lambda_2^{S(c\bar q)}$. It is known that the mass splitting of the pseudoscalar doublet originates from the chromomagnetic interaction \cite{Amoros:1997rx}
\be
{\lambda_2^{H(b\bar q)}\over \lambda_2^{H(c\bar q)} }={ m_{B^*}^2-m_B^2\over m_{D^*}^2-m_D^2}=\left( {\alpha_s(m_b)\over \alpha_s(m_c)}\right)^{9/25}\left[1-{\cal O}\left({\alpha_s\over \pi}\right)\right]+\Lambda_R\left({1\over m_c}-{1\over m_b}\right),
\en
where the non-perturbative parameter $\Lambda_R$ accounts for higher-order corrections in the heavy quark expansion. The next-to-leading correction for negative-parity mesons has been calculated in \cite{Amoros:1997rx}.
The experimental value of this ratio is $0.89\pm0.01$, while the leading logarithmic approximation gives 0.82. This means that the deviation of this ratio from unity is predominated by QCD corrections and the $1/m_Q$ effect is small. By the same token, we shall only consider the leading QCD correction to the relation $\lambda_2^{S(b\bar q)}=\lambda_2^{S(c\bar q)}$
\be \label{eq:lambda2corr}
\lambda_{2}^{S(b\bar q)}=\lambda_{2}^{S(c\bar q)}\left( {\alpha_s(m_b)\over \alpha_s(m_c)}\right)^{9/25},
\en
and a similar relation with the replacement of $\bar q$ by $\bar s$.
We then obtain
\be
\lambda_2^{S(b\bar q)}=(352\pm69\,{\rm MeV})^2, \qquad \lambda_2^{S(b\bar s)}=(372.5\pm1.2\,{\rm MeV})^2,
\en
and the masses (in MeV)
\be \label{eq:BmassQCD}
M_{B_{0}^*}=5715\pm22, \quad M_{B_{s0}^*}= 5715\pm1, \quad M_{B'_{1}}= 5752\pm31, \quad M_{B'_{s1}}= 5763\pm1.
\en
Comparing with Eq. (\ref{eq:BmassHQS}) we see that QCD corrections will mainly lower the masses of $B_0^*$ and $B_{s0}^*$ by an amount of 7 MeV.

We next discuss the corrections to $\Delta_S^{(b)}=\Delta_S^{(c)}$. In heavy quark effective theory, the mass of the heavy hadron $H_Q$ is of the form
\be \label{eq:HQET}
m_{H_Q}=m_Q+\bar\Lambda_{H_Q}-{\lambda_1\over 2m_Q}-{d_H\lambda_2\over 2m_Q},
\en
where the three nonperturbative HQET parameters $\bar\Lambda_{H_Q}$, $\lambda_1$ and $\lambda_2$ are independent of the heavy
quark mass,
\be
\lambda_1={\la H_Q|\bar Q(iD_\perp)^2Q|H_Q\ra\over 2m_{H_Q}}, \qquad
d_H\lambda_2={\la H_Q|\bar Q{1\over 2}\sigma\cdot G Q|H_Q\ra\over 2m_{H_Q}},
\en
with $d_H=-3$ for $J=0$ mesons and $d_H=1$ for $J=1$ mesons. Roughly speaking, $\bar\Lambda$ is the energy of light quark fields (i.e. brown muck), the $\lambda_1$ term represents the kinetic energy of the heavy quark $Q$ and $\lambda_2$ the chromomagnetic interaction energy. It follows that
\be
\la M_S\ra-\la M_H\ra=\bar\Lambda_S-\bar\Lambda_H-{\lambda_1^S\over 2m_Q}+{\lambda_1^H\over 2m_Q},
\en
and
\be \label{eq:1/mQcorr}
\Delta_S^{(b)}=\Delta_S^{(c)}+\delta\Delta_S\equiv \Delta_S^{(c)}+(\lambda_1^S-\lambda_1^H)\left( {1\over 2m_c}-{1\over 2m_b}\right).
\en
For the parameter $\lambda_1^H$, it can be determined from the relation
\be
\lambda_1^H=2m_bm_c\left( {\la M_H\ra^{(b\bar q)}-\la M_H\ra^{(c\bar q)}\over m_b-m_c}-1\right).
\en
This yields $\lambda_1^H=-0.13\,{\rm GeV}^2$ for $m_b=4.65$ GeV and $m_c=1.275$ GeV \cite{PDG} and is consistent with the value of $\lambda_1^H=-0.20\pm 0.06\,{\rm GeV}^2$ extracted from  the global fit \cite{Leibovich}. \footnote{If the $\ov{\rm MS}$ mass 4.18 GeV for $m_b$ is employed, we will have
$\lambda_1^H=1.6\,{\rm GeV}^2$.}
The parameter $\lambda_1^S$ is unknown. Nevertheless, we can learn something about it from another parameter $\lambda_1^T$ \cite{Mehen}. Denote the positive-parity spin doublet $(P_1,P_2^*)$ with $j^P={3\over 2}^+$ by the $T$ field so that its spin-averaged mass reads
\be
\la M_T\ra={5M_{P_2}+3M_{P_1}\over 8}.
\en
From the relation
\be
\lambda_1^T-\lambda_1^H=2m_bm_c\left( {\la M_T\ra^{(b\bar q)}-\la M_H\ra^{(b\bar q)}-\la M_T\ra^{(c\bar q)}+\la M_H\ra^{(c\bar q)}\over m_b-m_c}-1\right),
\en
and the measurement of masses for $P^*_2$ and $P_1$ \cite{PDG}, we find $\lambda_1^T-\lambda_1^H=-0.13\,{\rm GeV}^2$ for $q=u,d$ and $-0.17\,{\rm GeV}^2$ for $q=s$. This means that the magnitude of the kinetic energy of the excited heavy meson is larger than that in the ground states. Following \cite{Mehen}, it will be plausible to assume that the kinetic energy of the heavy quark in the $j^P={1\over 2}^+$ states is comparable to that of $j^P={3\over 2}^+$ states, say, $\lambda_1^S-\lambda_1^H\sim -0.12\,{\rm GeV}^2$. From Eq. (\ref{eq:1/mQcorr})
the $1/m_Q$ correction is estimated to be \footnote{The $1/m_Q$ correction $\delta\Delta_S$ was estimated to be $-50\pm25$ MeV in \cite{Mehen}.}
\be
\delta\Delta_S\sim {\cal O}(-35)\,{\rm MeV}.
\en
It is easily seen that as long as $\delta\Delta_S$ is not large in magnitude, the $1/m_Q$ correction to the relation $\Delta _S^{(b)}=\Delta _S^{(c)}$ [see Eq. (\ref{eq:1/mQcorr})], to a very good approximation, amounts to lowering the masses of $B^*_{(s)0}$ and $B'_{(s)1}$ by an equal amount of $|\delta\Delta_S|$:
\be \label{eq:Bmass1/mQ}
&& M_{B_{0}^*}=5715\pm22\,{\rm MeV}+\delta\Delta_S, \quad M_{B_{s0}^*}= 5715\pm1\,{\rm MeV}+\delta\Delta_S, \non \\
&& M_{B'_{1}}= 5752\pm31\,{\rm MeV}+\delta\Delta_S, \quad M_{B'_{s1}}= 5763\pm1\,{\rm MeV}+\delta\Delta_S.
\en
Evidently, the mass degeneracy between $B_{s0}^*$ and $B_0^*$ shown in Eq. (\ref{eq:BmassHQS}) implied by HQS is not spoiled by $1/m_Q$ and QCD corrections.

Finally, three remarks are in order: (i) The mass splitting of the scalar doublet in strange and nonstrange sectors obeys the relation
\be
{M_{B'_{s1}}-M_{B_{s0}^*} \over M_{B'_{1}}-M_{B_{0}^*} }={\lambda_2^{S(b\bar s)}\over \lambda_2^{S(b\bar q)} },
\en
where use of Eq. (\ref{eq:HQET}) has been made. From (\ref{eq:lambda2corr}), (\ref{eq:BmassQCD}) or (\ref{eq:Bmass1/mQ}), one can check that the above relation is numerically respected. (ii) It is interesting to notice that the relation $M_{B'_{s1}}-M_{B_{s0}^*}\simeq M_{B^*}-M_B=45.78\pm0.35$ MeV is expected in the scenario where $B_{s0}^*$ and $B'_{s1}$ are the $BK$ and $B^*K$ bound states, respectively \cite{Cleven:2010aw}. This relation is indeed confirmed as we have the mass difference $M_{B'_{s1}}-M_{B_{s0}^*}=48\pm1$ MeV from Eq. (\ref{eq:BmassQCD}) or (\ref{eq:Bmass1/mQ}). However, here we do not rely on the bound-state assumption.
(iii) Since the masses of $B_{s0}^*$ and $B'_{s1}$ are below the $BK$ and $B^*K$ thresholds, respectively, their widths are expected to be very narrow with the isospin-violating strong decays into $B_s\pi^0$ and $B_s^*\pi^0$.

\section{Near degeneracy between $K_0^*(1430)$ and $a_0(1450)$}
As discussed in the Introduction, there have been several
attempts to understand the near degeneracy between $a_0(1450)$ and
$K^*_0(1430)$. (i)  In the conventional QCD sum rule, $K_0^*(1430)$ is heavier than $a_0(1450)$. By introducing instanton effects to the correlation function, it can be the other way around \cite{Jin:light}. (ii) In analog to the interpretation of the light scalar mass spectrum in terms of the tetraquark picture, it is tempting to assume that heavy scalar mesons $f_0(1370)$, $K_0^*(1430)$, $a_0(1450)$ and $f_0(1500)$
form another tetraquark nonet in order to explain the reversed ordering of the observed spectrum \cite{Maiani}. This assumption encounters two difficulties. First, another scalar $q\bar q$ nonet around 1.2 GeV should exist, but it has not been observed. Second, the quenched lattice
calculations in \cite{Mathur} with the bare
$q\bar{q}$ states already
suggests the near degeneracy between $a_0(1450)$ and
$K^*_0(1430)$.
(iii) The mixing of
the $q\bar q$ heavy scalar nonet with the light tetraquark nonet will push up the masses of $K_0^*(1430)$ and $a_0(1450)$ and push down the masses of $K_0^*(800)$ and $a_0(980)$ to render the former two scalars closer \cite{Schechter} (see, however, a criticism from \cite{Maiani}).  It was proposed in \cite{Hooft} that this mixing can be generated by instanton-induced interactions.

As the mass shift due to self-energies is more effective in the strange sector than the non-strange one for heavy scalar mesons, it is natural to consider the self-energy corrections to $K_0^*(1430)$ and $a_0(1450)$. The strong decay of the former is the $K\pi$ mode, while the hadronic decay modes $\pi\eta,\pi\eta',K\ov K$, $\omega\pi\pi$ and $a_0(980)\pi\pi$ of $a_0$ have been measured. It appears that the strong decay of $a_0(1450)$ is dominated by $\omega\rho$ and $a_0(980)f_0(500)$ \cite{Bugg}.

We shall consider the derivative couplings of a scalar meson to two light pseudoscalar mesons given in Eq. (\ref{eq:SPPcoupling}).
The physical mass $M^2_{K_0^*}$ of $K_0^*(1430)$ is given by the equation
\be
&& M^2_{K_0^*}-{\cal M}^2_{K_0^*}-{3\over 2}{\rm Re}\,\Pi_{K\pi}(M^2_{K_0^*})=0,
\en
while the width is
\be \label{eq:K0stwidth}
\Gamma=-{3\over 2}{1\over M_{K_0^*}}Z\,{\rm Im}\,\Pi_{K\pi}(M_{K_0^*}^2),
\en
where $\Pi_{K\pi}(s)$ has the same expression as Eq. (\ref{eq:loops}) except that the term $(2g^2/f_\pi^2)$ is replaced by $g_{K_0^*(1430)\to K^+\pi^-}^2$.
We obtain
\be \label{eq:g}
g_{K_0^*(1430)\to K^+\pi^-}=4.21\pm0.64\,{\rm GeV}^{-1}
\en
from the experimental measurement $\Gamma(K_0^*)=270\pm80\,{\rm MeV}$ \cite{PDG}. The renormalization scale $\Lambda$ is fixed in such a way that the calculated $K_0^*$ width through the imaginary part of the chiral loop via Eq. (\ref{eq:K0stwidth}) agrees with experiment.

We see from Table \ref{tab:K0st} that self-energy corrections will push down the mass of $K_0^*$ by an amount depending on the input bare mass. At this moment, we cannot say anything for the $a_0(1450)$ as we haven not studied chiral loop effects due to the intermediate states $\omega\rho$ and $a_0(980)f_0(500)$.  (The decay modes with two pseudoscalar mesons $\pi\eta,\pi\eta',K\ov K$ are far from threshold.) It has been conjectured that such effect may even push up the mass of $a_0(1450)$ \cite{Bugg}.

\begin{table}[t]
\caption{Mass shift  $\Delta M_{K_0^*}$ of the $K_0^*(1430)$ meson due to self-energy loop effects. The renormalization scale $\Lambda$ is fixed in such a way that the imaginary part of the chiral loop will produce the measured $K_0^*$ width. For illustration we consider two values for the bare mass ${\cal M}$ of $K_0^*(1430)$.
All the scales  are in MeV.} \label{tab:K0st}
\begin{tabular}{ | c c c c  |}  \hline
 ${\cal M}_{K_0^*}$ & $\Lambda$ & $\Delta M_{K_0^*}$ & $M_{K_0^*}$  \\ \hline
 ~~~1500~~~ & 1150 & ~~$-36$~~ & ~~1464~~ \\
 1550 & 1500 & $-62$ & 1489 \\ \hline
\end{tabular}
\end{table}

Finally we remark that even in the SU(3) limit, not all the ten states $f_0(1370),f_0(1500),f_0(1710)$, $a_0(1450),K_0^*(1430)$ are degenerate in masses. This is because of the quark-antiquark annihilation and glueball contributions to the isosinglet scalar mesons. It has been shown in \cite{CCL} that in the SU(3) limit, $f_0(1500),a_0(1450)$ and $K_0^*(1430)$ are degenerate and $f_0(1500)$ is an SU(3) octet state $|f_{\rm octet}\ra= {1\over\sqrt{6}}(|u\bar u\ra+|d\bar d\ra-2|s\bar  s\ra)$ so that it does not receive quark-antiquark annihilation contribution. In the absence of
glueball-quarkonium mixing, $f_0(1370)$ becomes a pure SU(3) singlet
$|f_{\rm singlet}\ra={1\over\sqrt{3}}(|u\bar u\ra+|d\bar d\ra+|s\bar  s\ra)$ and $f_0(1710)$ a pure glueball $|G\ra$. The state $f_0(1370)$ is lighter than $f_0(1500)$ due to the negative contributions from quark-antiquark annihilation. When glueball-quarkonium mixing is turned on, $f_0(1370)$ is mainly an SU(3) singlet with a slight mixing with the scalar glueball which is the primary component of $f_0(1710)$.

\section{Conclusions}

In this work we have considered
the empirical observation of near degeneracy of scalar mesons above 1 GeV, namely, the mass of the strange-flavor scalar meson is similar to that of the non-strange one. Lattice calculations of the
$a_0(1450)$ and $K_0^*(1430)$ masses demonstrate that, to first order approximation, flavor SU(3) is a good
symmetry for the scalar mesons above 1 GeV.
Many studies indicate that the low mass of $D_{s0}^*(2317)$ ($D_0^*(2400)$) arises from the mixing between the $0^+$ $c\bar s$ ($c\bar q$) state and the $DK$ threshold ($D\pi$ state). This has been realized in QCD sum rule and lattice calculations. It is thus important to check if
the mass difference between strange heavy scalar mesons and their non-strange partners can be substantially reduced by self-energy contributions.

\begin{itemize}

\item
We first work in HMChPT. Contrary to the common practice, the renormalization scale has to be larger than the chiral symmetry breaking scale of order 1 GeV in order to satisfy the on-shell conditions.
Self-energy corrections in general will push down the masses of $B_{s0}^*$ much more than that of $B_0^*$, but this is not the case for $D_{s0}^*$ and $D_0^*$.
Moreover,
near mass degeneracy and the physical masses of $D_{s0}^*$ and $D_0^*$ cannot be accounted for simultaneously in this approach.

\item
The conventional model without heavy quark expansion  works better than HMChPT. The approximate mass degeneracy can be qualitatively understood as a consequence of self-energy hadronic loop corrections which will push down the mass of the heavy scalar meson in the strange sector more than that in the non-strange one. However, quantitative predictions on the chiral loop calculations are subject to many theoretical uncertainties, especially the loop results are sensitive to the choice of the parameter $\Lambda$, which is an arbitrary renormalization scale. Physics should be independent of the $\Lambda$, but this issue cannot be properly addressed in the phenomenological model discussed here.  Nevertheless, the different choices of $\Lambda$ in the first model based on HMChPT and in the second model with the assumption that at some substraction scale the contribution from any hadronic loop vanishes, could be the reason why the latter model works empirically better than the former.

\item
In the heavy quark limit, near mass degeneracy observed in the scalar charm sector will imply the same phenomenon in the $B$ system.  The predictions $M_{B_0^*}\approx M_{B_{s0}^*}\approx 5715\,{\rm MeV}+\delta\Delta_S$, $M_{B'_1}\approx 5752\,{\rm MeV}+\delta\Delta_S$  and $M_{B'_{s1}}\approx 5763\,{\rm MeV}+\delta\Delta_S$ are based on heavy quark symmetry and the leading QCD correction, where $\delta\Delta_S$ stems from $1/m_Q$ corrections. A crude estimate yields $\delta\Delta_S\sim {\cal O}(-35\,{\rm MeV})$  or  less. We stress that the closeness of $B_{s0}^*$ and $B_0^*$ masses implied by HQS is not spoiled by $1/m_Q$ or QCD corrections.  Measurements of these scalar $B$ states will test if they are degenerate in masses and if their masses lie in the range of $5680-5715$ MeV.

\item
The mass-shift effect on $K_0^*(1430)$ will lower its mass. However, we cannot say anything for the $a_0(1450)$ at this moment as we have not studied chiral loop effects due to the intermediate channels $\omega\rho$ and $a_0(980)f_0(500)$, though it has been conjectured that such effect may even push up the $a_0(1450)$ mass.

\end{itemize}

In short, the similarity of the mass of the strange-flavor heavy scalar meson with that of its non-strange partner can be qualitatively understood as a consequence of self-energy effects due to strong coupled channels. Quantitatively, the masses of $B_{s0}^*$ and $B_0^*$ and their degeneracy can be deduced from the charm spectroscopy and heavy quark symmetry with corrections from QCD and $1/m_Q$ effects.

\vskip 5.0cm \acknowledgments
This research was supported in part by the Ministry of Science and Technology of R.O.C. under the Grant No. 100-2112-M-001-009-MY3 and by the National Science Foundation of China under the Grant No. 11347027.


\end{document}